\documentstyle[psfig]{lamuphys}
\makeatletter
\let\chapter\hid@chapter
\makeatother

\newcounter{bibcnt}
\newcommand{\bibi}[1]{\bibitem{[\thebibcnt]}{#1}{[\thebibcnt]}\stepcounter{bibcnt}}
\begin{document}
\title{Shot-noise 
in non-degenerate semiconductors with energy-dependent elastic scattering}
\titlerunning{Shot-noise in non-degenerate semiconductors}
\author{H. Schomerus\inst{1},
E. G. Mishchenko\inst{1,2},
 and C. W. J. Beenakker\inst{1}}
\institute{Instituut-Lorentz, Universiteit Leiden,
P.\,O.~Box 9506, 2300 RA Leiden,
  The Netherlands
\and
L. D. Landau Institute for Theoretical Physics, Kosygin 2, Moscow
117334, Russia
}
\date{(July 1999)}
\maketitle

\begin{abstract}
We investigate current fluctuations in non-degenerate
semiconductors, on length scales intermediate between the
elastic and inelastic mean free paths.
The shot-noise power $P$ is suppressed 
below the Poisson value $P_{\rm Poisson}=2e\bar I$
(at mean current $\bar I$) by the Coulomb repulsion
of the carriers.
We consider a
power-law dependence of the elastic scattering time
$\tau\propto \varepsilon^{\alpha}$
on kinetic energy $\varepsilon$ and
present an exact solution of the non-linear
kinetic equations in the regime of space-charge
limited conduction. The ratio $P/P_{\rm Poisson}$ decreases from 
$0.38$ to $0$ in the range $-\frac{1}{2}< \alpha< 1$.
\end{abstract}

\section{Introduction}

The noise power $P$ of current fluctuations in an electron gas in
thermal equilibrium (at temperature $T$) is related by the
Johnson-Nyquist formula $P=4kTG$ (with $k$ Boltzmann's constant)
to the linear-response conductance $G=\lim_{V\to 0} d\bar I/ dV$
(with $\bar I$ the mean current in response to an applied voltage $V$).
This formula can be generalized to a large applied voltage,
$P=4kT(V/\bar I)(d\bar I/dV)^2$, provided the electron gas remains in
local equilibrium with the lattice.
Local equilibrium requires inelastic scattering. When the conductor is
shorter than the inelastic mean free path $l_{\rm in}$ and the potential
drop $V$ is large enough,
the Johnson-Nyquist formula no longer
applies and a measurement of current noise (then also called shot noise) 
reveals more detailed
information about the transport of charge carriers---in
particular about their correlations. 
The maximal noise level $P_{\rm Poisson}=2e\bar I$
is attained in absence of all
correlations (both in the injection process as well as in the subsequent
transport). Examples are vacuum diodes at large bias
in absence of space-charge effects and tunneling diodes with low
transmissivity.

Here we consider the transport through a disordered semiconductor
of length $L$
terminated by two metal contacts,
under the conditions of elastic scattering
($l\ll L\ll l_{\rm in}$,  with $l$ 
the elastic mean free path).
In a degenerate conductor correlations are
induced by the Pauli exclusion principle (for a review 
of the theory of shot noise in this situation see Ref.~\cite{review})
and the shot noise has the universal value
$P=\frac{1}{3} P_{\rm Poisson}$ \cite{Buettiker,Nagaev}.

At low carrier concentration the electron gas is non-degenerate,
and the Pauli principle is ineffective. Because carriers can
now accumulate, giving rise to space-charge effects,
they become correlated
through Coulomb repulsion. This is the situation which we want to
study presently.
In a recent Monte-Carlo simulation \cite{Gonzalez} 
a shot-noise suppression factor
of about $P/P_{\rm Poisson}=1/3$ was found 
in the regime of space-charge limited transport;
an energy-independent elastic scattering rate was assumed.
The coincidence with the noise level
obtained in the degenerate situation 
attracted a lot of attention \cite{Landauer}. 
The degree of universality
is less pronounced here since the number actually depends on the
geometry and dimensionality---as
well as the scattering mechanism
\cite{Beenakker1998,schomerus1999,Nagaev1998}.

In Ref.~\cite{Beenakker1998} the problem 
was investigated for an energy-independent elastic
scattering time $\tau$, using the kinetic theory of non-equilibrium 
fluctuations (reviewed in Ref.~\cite{Book}).
The non-linear kinetic equations were solved in a certain approximation
(the drift approximation), with the result $P/P_{\rm Poisson}=0.3410$.
In Ref.~\cite{schomerus1999} we obtained an exact solution,
giving $P/P_{\rm Poisson}=0.3097$, and also considered 
a power-law dependence $\tau\sim\varepsilon^\alpha$ on the kinetic energy
$\varepsilon$.
For $\alpha=-\frac{1}{2}$ (corresponding to short-range impurity
scattering or quasi-elastic acoustic phonon scattering \cite{gantsevich1979})
we found the exact result $P/P_{\rm Poisson}=0.3777$.
For other values of $\alpha$
we only presented results within the drift approximation. 
In this work we derive the exact solution in the range $-\frac{1}{2}<\alpha
< 1$.
As we will discuss, $\alpha$ should be in this range for
space-charge limited conduction to be realized.

\section{The drift-diffusion equation}

We consider a three-dimensional conductor of length $L$ and
cross-sectional area $A$ terminated by
two contacts.
The equilibrium density $\rho_{\rm eq}$ of charge
carriers (charge $e$, effective mass $m$) in the decoupled
conductor is assumed to be much lower than 
the density $\rho_c$ of those carriers that are energetically
allowed (at a given voltage $V$)
to enter the conductor from the contacts.
(A possible realization would be an
intrinsic or barely doped
semiconductor between two metal contacts or two heavily doped
semiconducting regions.) 
The dielectric constant of the conductor is $\kappa$.
The temperature $T$ is assumed to be so high that the electron gas 
is degenerate, and a large voltage drop $V\gg kT/e$
is maintained between the contacts.
Transport is assumed to be diffusive and elastic,
$l<L<l_{\rm in}$. We assume a power-law energy dependence
\begin{equation}
\tau(\varepsilon)=\tau_0\varepsilon^\alpha
\label{eq:tdep}
\end{equation}
of the elastic scattering time on the kinetic energy $\varepsilon$.
We want to calculate the zero-frequency component
\begin{equation}
P=2\int_{-\infty}^\infty{\rm d}t'\,\overline{\delta I(t) \delta I(t+t')}
\label{eq:powerdef}
\end{equation}
of the noise power of the fluctuations $\delta I(t)$ of the electric
current $I(t)=\bar I+\delta I(t)$ around its mean $\bar I$.

We use Cartesian coordinates $x,y,z$ with $x$ parallel to the conductor
(the current source is at $x=0$, the drain at $x=L$).
To linear order in the fluctuations, the transverse coordinates can
be ignored. In the zero-frequency limit the current is independent
on $x$ because of the continuity equation and
is given by the drift-diffusion equation
\cite{Beenakker1998,schomerus1999}
\begin{equation}
\label{eq:dd2new}
I(t)=-\frac{\partial}{\partial x}
\int d\varepsilon \,
D(\varepsilon)\rho(x,\varepsilon,t)
+E(x,t)
\int d\varepsilon \,
{\cal F}(x,\varepsilon,t)
\frac{d\sigma(\varepsilon)}{d\varepsilon}
+\delta J(x,t) 
.
\label{eq:ddx}
\end{equation}
The electric field $E(x,t)$ is related to
the laterally integrated charge density $\rho(x,t)$ by the Poisson
equation
\begin{equation}
\label{eq:poisson2}
\kappa  \frac{\partial}{\partial x}E(x,t)
=\frac{1}{A} \rho(x,t)
,
\end{equation}
where we omitted the low background charge density $-\rho_{\rm eq}$.
The fluctuating source $\delta J(x,\varepsilon,t)$ accounts
for the stochasticity of individual scattering events and has the
correlator
\begin{equation}
\label{eq:correl3}
\overline{\delta {J}(x,t)
\delta  {J}(x',t') }
 =
2 A \delta(t-t')\delta(x-x')
\int{\rm d}\varepsilon\,
\sigma(\varepsilon)\bar {\cal F}(x,\varepsilon)
 .
\end{equation}
Here and in Eq.~(\ref{eq:ddx}), ${\cal F}(x,\varepsilon,t)=\rho(x
,\varepsilon,t)/e\nu(\varepsilon)$ with the density of states
$ \nu(\varepsilon)=
4 \pi m (2m\varepsilon)^{1/2}=\nu_0 \varepsilon^{1/2}
$
(we set Planck's constant $h\equiv 1$).
The conductivity
$ \sigma(\varepsilon)=e^2\nu(\varepsilon) D(\varepsilon)
=\sigma_0\varepsilon^{\alpha+3/2}
$
is the product
of the
density of states and the
diffusion constant
$ D(\varepsilon)= v^2 \tau/3=D_0 \varepsilon^{\alpha+1}$.

\section{Space-charge limited conduction}
\label{sec:scl}
For a large voltage drop $V$ between the two metal contacts
and a high carrier density $\rho_c$ in the contacts,
the charge injected into the semiconductor is much higher than the
equilibrium charge $\rho_{\rm eq}$, which can then be neglected.
For sufficiently high $V$ and $\rho_c$ the system
enters the regime of space-charge limited conduction \cite{Lampert},
defined by the boundary condition 
\begin{equation}
\label{eq:bc1a}
E(x,t)=0\quad\mbox{at}\quad x=0 
.
\end{equation}
Eq.~(\ref{eq:bc1a}) states that the space charge 
$Q=\int_0^L\rho(x)\,{\rm d}x$ in the semiconductor is precisely
balanced by the surface charge at the current drain.
At the drain we have the absorbing boundary condition
\begin{equation}
\label{eq:bc2a}
\rho(x,t)=0\quad\mbox{at} \quad x=L 
.
\end{equation}
With this boundary condition we again 
neglect $\rho_{\rm eq}$.

To determine the electric field inside the semiconductor we proceed as
follows.
Since scattering is elastic, the total energy
$u=\varepsilon-e\phi(x,t)$ of each carrier is preserved.
The potential gain $-e\phi(x,t)$ (with $E=-\partial
\phi/\partial x$) dominates over the initial thermal
excitation energy of order $kT$
almost throughout the whole
semiconductor; only close
to the current source
(in a thin boundary layer)
this is not the case.
We can therefore approximate the kinetic energy $\varepsilon\approx-e\phi$
and introduce 
this into $D(\varepsilon)$ and $d\sigma/d\varepsilon$.
Substituting
into Eq.~(\ref{eq:dd2new})
one obtains
\begin{eqnarray}
{\cal F}(x,t)
&\approx&
e \int_x^Ldx'\,
\frac{I(t)-\delta J(x,t)}{
\sigma_0[-e\phi(x',t)]^{\alpha+3/2}}
,
\label{eq:f}
\\
\rho(x,t)&\approx&
\frac{[-e\phi(x,t)]^{1/2}}{D_0} \int_x^Ldx'\,
\frac{I(t)-\delta J(x,t)}{
[-e\phi(x',t)]^{\alpha+3/2}}
,
\nonumber\\
\label{eq:barf}
\end{eqnarray}
where the absorbing boundary conditions have been used.
From the Poisson equation (\ref{eq:poisson2})
we find the third-order, non-linear, inhomogeneous
differential equation
\begin{eqnarray}
\label{eq:dgl}
&&2(-\phi)^\alpha
\phi'
\phi''
+4 
(-\phi)^{\alpha+1}
\phi'''
=
B\bar I [1+\delta i (x,t)]
,
\\
&&\delta i(x,t) = \frac{I(t)-\delta J(x,t)}{\bar I}
,
\end{eqnarray}
for the potential profile $\phi(x,t)$.
Primes denote differentiation with respect to $x$, and
$B=6/e^\alpha\mu_0\kappa A$ with
$\mu_0=e\tau_0/m$.

Since the potential difference $V$ between source and drain does not
fluctuate, we have the two boundary conditions
$\phi(0,t)=0$, $\phi(L,t) =- V$.
Eqs.~(\ref{eq:bc1a}) and (\ref{eq:bc2a}) imply
two additional boundary conditions,
$\phi'(0,t)=0$,
$\phi''(L,t)=0$.

The differential equation  (\ref{eq:dgl}) and the
accompanying boundary conditions
possess two remarkable scaling properties: The product $B\bar I$
of material parameters and mean current $\bar I$ and the
length $L$ can be eliminated by introduction of the scaled potential
\begin{equation}
\chi(x,t)=-\left(L^3 B\bar I\right)^{-1/(\alpha+2)}\phi(xL,t)
.
\end{equation}
The rescaled differential equation reads
\begin{equation}
2\chi^\alpha \chi' \chi'' -4 \chi^{\alpha+1} \chi''' = 1+\delta i
,
\label{eq:dgl2}
\end{equation}
which has to be solved with the boundary conditions
$\chi(0,t)=0$,
$\chi(1,t) = \left(L^3 B\bar I\right)^{-1/(\alpha+2)}V$,
$\chi'(0,t)=0$,
$\chi''(1,t)=0$.
The scaling properties entail that
the shot-noise suppression factor depends only on the exponent $\alpha$,
but no
longer
on the parameters
$L$, $A$, $V$, $\tau_0$, and $\kappa$.

We will solve this boundary value problem for $\chi=\bar \chi+\delta\chi$,
first for the mean (Section \ref{sec:mean})
and then for the fluctuations (Section \ref{sec:fluctuations}), in both cases
neglecting terms quadratic in $\delta\chi$.

\section{Average profiles and the current-voltage characteristic}
\label{sec:mean}

The averaged equation (\ref{eq:dgl2}) for the rescaled mean potential $\bar
\chi(x)$ reads
\begin{equation}
\label{eq:mean}
2\bar \chi^\alpha \bar \chi' \bar \chi'' -4 \bar \chi^{\alpha+1} \bar \chi'''
= 1
.
\end{equation}
We seek a solution
which fulfills the three boundary
conditions $\bar \chi(0)=0$, $\bar \chi'(0)=0$, $\bar \chi''(1)=0$.
The value of $\bar\chi$ at the current drain
determines the current-voltage characteristic
\begin{equation}
\label{eq:iv}
\bar I(V)=\frac{1}{L^3B} \left(\frac{V}{\bar \chi(1)}\right)^{\alpha+2}
.
\end{equation}
We now construct $\bar \chi(x)$.
The function
$
\bar \chi_0(x)= a_0 x^\beta$ with
$\beta=3/(2+\alpha)$ and
$a_0=\left[2\beta(\beta-1)(4-\beta)\right]^{-\beta/3}
$
solves the differential equation and
satisfies the boundary conditions at $x=0$, but 
$\bar \chi''_0(x)\neq 0$ for any finite $x$.
We substitute into Eq.~(\ref{eq:mean}) the ansatz 
$
\bar \chi(x)=\sum_{l=0}^\infty
a_l x^{\gamma l+\beta}
$,
consisting of $\bar \chi_0(x)$
times a power series in $x^\gamma$, with $\gamma$ a
positive power to be determined.
This ansatz proves fruitful since
both terms on the left-hand side of Eq.~(\ref{eq:mean}) give the same powers
of $x$, starting with order $x^0$ in coincidence with the right-hand side.
By power matching one obtains in first order the value for $a_0$
given above.
The second order leaves $a_1$ as a
free coefficient, but fixes the power 
$
\gamma=(8-5\beta+\sqrt{-32+40\beta+\beta^2})/4$
.
The
coefficients
$a_l$ for $l\ge 2$ are then given recursively as a function of
$a_1$, which is finally determined from
the condition $\bar \chi''(1)=0$.

\begin{figure}[thbp]
\psfig{file=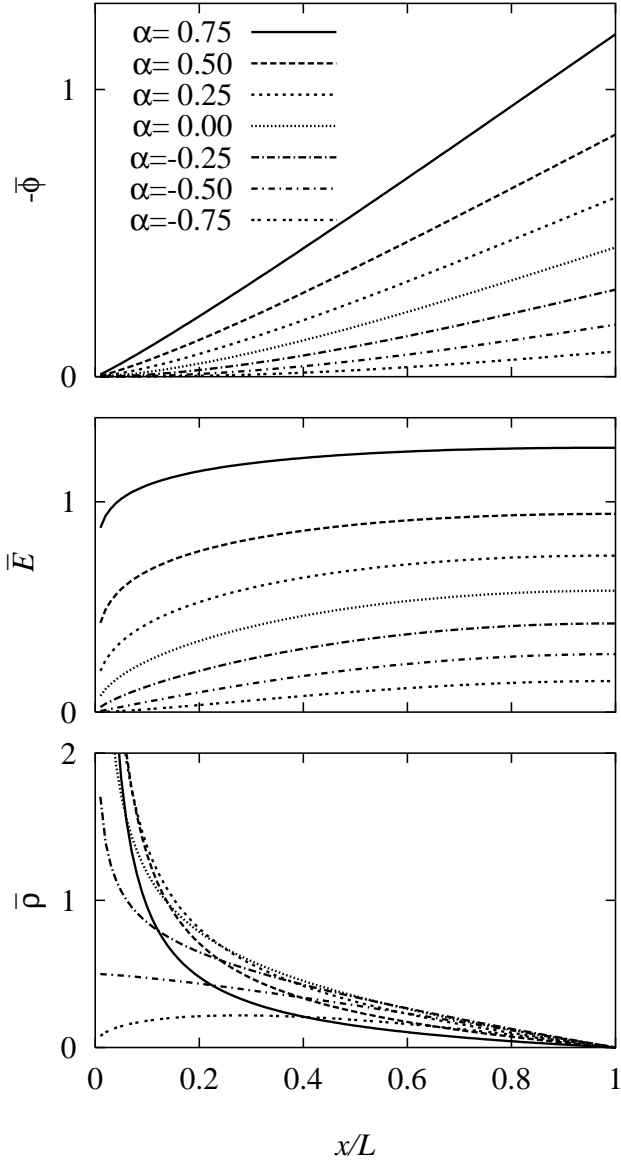,width=9cm,clip=}
\caption[]{
Profile of the mean electrical potential $\bar \phi$ [in units of
$(L^3B\bar I)^{1/(\alpha+2)}$, with
$B=6m/e^{\alpha+1}\tau_0\kappa A$], 
the electric field $\bar E$ [in units of $(L^3B\bar I)^{1/(\alpha+2)}/L$],
and the charge density $\bar \rho$
[in units of $\kappa (L^3B\bar I)^{1/(\alpha+2)}/L^2$], following from
Eq.~(\protect\ref{eq:mean}) for different values of $\alpha$.
}
\label{fig:fig1}
\end{figure}
In Fig.~\ref{fig:fig1}
the profiles
of the potential $\bar\phi\propto\bar  \chi$, the electric field $\bar E\propto
\bar \chi'$, and the charge density
$\bar\rho\propto\bar \chi''$ are plotted for various values of $\alpha$.
The coefficient 
$\bar \chi(1)$ appearing in the current-voltage characteristic
(\ref{eq:iv}) can be read off from this plot.
The behavior at the current source changes qualitatively at
$\alpha=-\frac{1}{2}$ (see Section \ref{sec:concl}).

\section{Fluctuations}
\label{sec:fluctuations}

The rescaled fluctuations $\delta\chi(x,t)=\psi(x,t)$
fulfill the linear differential equation
\begin{eqnarray}
{\cal L}[\psi]&=&-4\bar \chi^{\alpha+1}\psi'''
+2\bar \chi^{\alpha}\bar \chi'\psi''
+2\bar \chi^{\alpha}\bar \chi''\psi'
\nonumber
\\&&{}
+
\left[2\alpha\bar \chi^{\alpha-1}
\bar \chi'\bar \chi''-4(\alpha+1)
\bar \chi^{\alpha}\bar \chi'''
\right]\psi
=\delta i(t)
.
\end{eqnarray}
The solution of the inhomogeneous equation
is found with help of the three independent 
solutions
of the homogeneous equation ${\cal L}[\psi]=0$,
$ \psi_1(x)=\bar \chi'(x)$,
$ \psi_2(x)=\bar \chi(x)- (x/\beta) \bar \chi'(x)$, and
\begin{equation}
\psi_3(x)=
\psi_1(x)\int_x^1{\rm d}x'\,\frac{\bar \chi^{1/2}(x')\psi_2(x')}{
{\cal W}^2(x')}
-
\psi_2(x)\int_x^1{\rm d}x'\,\frac{\bar \chi^{1/2}(x')\psi_1(x')}{
{\cal W}^2(x')}
,
\end{equation}
where we have defined
$
{\cal W}(x)=\psi_1(x)\psi_2'(x)- \psi_1'(x)\psi_2(x)
$
.
The special solution which fulfills $\psi(0,t)=\psi'(0,t)=\psi(1,t)=0$
is
\begin{eqnarray}
\psi(x,t)&=&
\int_0^1{\rm d}x'\,\frac{\bar \chi^{1/2}(x')}{{\cal W}^2(x')}
\Big[\Theta(x-x')\psi_1(x)\psi_2(x')
+\Theta(x'-x)\psi_1(x')\psi_2(x)
\nonumber
\\
&&
{}
-\frac{\psi_1(1)}{\psi_2(1)}\psi_2(x)\psi_2(x')
\Big]
\int_0^{x'}{\rm d}x''\,
\frac{\delta I(t)-\delta J(x'',t)}{4\bar I }\frac{{\cal W}(x'')}{
\bar \chi^{\alpha+3/2}(x'')}
.
\end{eqnarray}

The condition $\psi''(1,t)=0$ relates the fluctuating current $\delta I$
to the Langevin current $\delta J$.
The resulting
expression is of the form
\begin{equation}
\delta I(t)=
{\cal C}^{-1}
\int_0^L{\rm d}x\,
\delta {J}(x,t){\cal G}(x)
 ,
\label{eq:deltai}
\end{equation}
with the definitions
$
{\cal C}=\int_0^1{\rm d}x\,{\cal G}(x)
$,
\begin{equation}
{\cal G}(x)=\frac{{\cal W}(x)}{\bar \chi^{\alpha+3/2}(x)}
\left(
1+\frac{(1-1/\beta)\bar \chi^{\prime 2}(1)}{
4\bar \chi^{\alpha+1/2}(1)\psi_2(1)}
\int_x^1{\rm d}x'\,
\frac{\bar \chi^{1/2}(x')\psi_2(x')}{{\cal W}^2(x')}
\right)
.
\label{eq:calgdep}
\end{equation}

The shot-noise power is found by substituting Eq.~(\ref{eq:deltai})
into Eq.~(\ref{eq:powerdef}) and invoking the correlator
(\ref{eq:correl3}) for the Langevin current.
This results in 
\begin{equation}
\label{eq:pres1}
P=2\int_0^L{\rm d}x\,\left(\frac{{\cal G}(x)}{\cal C}\right)^2
{\cal H}(x) 
\end{equation}
with
${\cal H}(x)=
2A\int{\rm d}\varepsilon\,\sigma(\varepsilon)
\bar{\cal F}(x,\varepsilon)
\approx 2\sigma_0[-e\bar\phi(x)]^{\alpha+3/2}\bar{\cal F}(x)
$.
Eq.~(\ref{eq:f}) gives
\begin{equation}
{\cal H}(x)=2e\bar I
\bar \chi^{\alpha+3/2}(x)\int_x^1 dx'\,\frac{1}{\bar \chi^{\alpha+3/2}(x')}
=
4P_{\rm Poisson}\bar \chi^{\alpha+1}(x)\bar \chi''(x)
,
\label{eq:calhdep}
\end{equation}
where we integrated with help of Eq.~(\ref{eq:mean}) and used
$\bar \chi''(1)=0$.

\begin{figure}[thbp]
\psfig{file=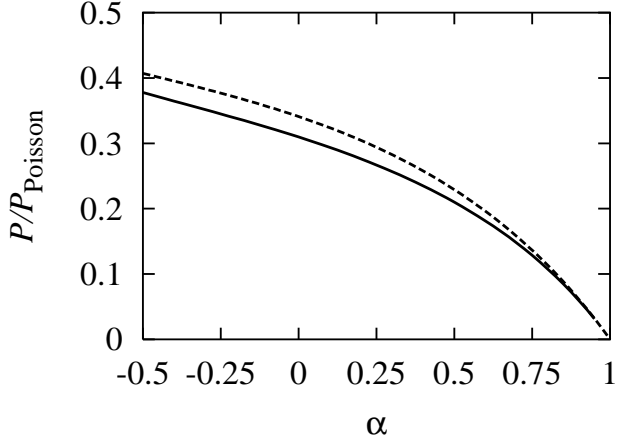,width=9cm,clip=}
\caption[]{
Shot-noise power $P$ 
as a function of $\alpha$. 
The exact result (solid curve) is compared with the approximate result
(\protect\ref{eq:driftapp}) (dashed curve).
}
\label{fig:fig2}
\end{figure}

In Fig.~\ref{fig:fig2} we plot the ratio $P/P_{\rm Poisson}$ as a function
of the parameter $\alpha$ (solid curve).
The shot-noise 
suppression factor $P/P_{\rm Poisson}= 0.3777$ for
$\alpha=-\frac{1}{2}$ and goes to zero as $\alpha\to 1$.

\section{Drift approximation}
A simple formula for the shot-noise suppression factor
can be found when one neglects the diffusion term in Eq.~(\ref{eq:ddx})
and considers instead of Eq.~(\ref{eq:dgl2})
the corresponding differential equation
$
(4\alpha+6)\chi^\alpha \chi' \chi'' = 1+\delta i
$
.
The is the drift approximation of Ref.~\cite{Beenakker1998}.
The order of the differential equation is reduced by one, so that we also
have to drop one of the boundary conditions.
The absorbing boundary condition $\chi''(1,t)=0$ is the most reasonable
candidate, because even for the resulting mean profile
$\bar \chi(x)=b_0x^\beta$ with
$\beta=3(2+\alpha)^{-1}$ and
$b_0=[\beta^2(\beta-1)]^{-\beta/3}$
most carriers remain
concentrated close to the current source.
The differential equation for the fluctuations
$\alpha{\psi}/{\bar \chi}+{\psi'}/{\bar \chi'}+
{\psi''}/{\bar\chi''} =\delta i$
can be solved with help of the homogeneous solutions
$
\psi_1(x)=x^{\beta-1}$ and
$\psi_2(x)=x^{3-2\beta}$.
The inhomogeneous solution that fulfills $\psi(0,t)=0$, $\psi'(0,t)=0$
is
\begin{eqnarray}
\psi(x,t)&=&b_0\frac{\beta(\beta-1)}{4-3\beta}\int_0^xdx'\,\left[
x^{3-2\beta}
x'^{3\beta-4}
-x^{\beta-1}
\right]\delta i(x',t)
.
\end{eqnarray}
We demand that the voltage does not fluctuate, $\psi(1,t)=0$, and obtain
Eq.~(\ref{eq:deltai}) with now
${\cal G}(x)=1-x^{3\beta-4}$.
The shot noise power is finally found from
Eq.~(\ref{eq:pres1}) with ${\cal H}(x)=P_{\rm Poisson}x^{3-\beta/2}
\int_x^1dx'\,x'^{\beta/2-3}$, 
\begin{eqnarray}
P/P_{\rm Poisson} &=&
\frac{6(\alpha-1)(\alpha+2)(16\alpha^2+36\alpha-157)}{
5(2\alpha-5)(8\alpha-17)(13+8\alpha)}
 .
 \label{eq:driftapp}
\end{eqnarray}
This is the dashed curve in Fig.~\ref{fig:fig2}.

\section{Discussion}
\label{sec:concl}

The shot-noise suppression factor $P/P_{\rm Poisson}$
varies from $0.38$ to $0$ in the range $-\frac{1}{2}< \alpha <1$,
which includes
the case of an energy-independent elastic scattering rate
($\alpha=0$, $P/P_{\rm Poisson}=0.3097$) and the case of short-range scattering
by uncharged impurities or  quasi-elastic scattering by acoustic
phonons
($\alpha=-\frac{1}{2}$, $P/P_{\rm Poisson}=0.3777$).
The results in the drift approximation (\ref{eq:driftapp})
are about $10\%$ larger.
Our values are somewhat smaller than those following from 
the numerical simulations 
of Gonz{\'a}lez {\it et al.}, who found
$P/P_{\rm Poisson}= \frac{1}{3}$ for $\alpha=0$ \cite{Gonzalez}
and $P/P_{\rm Poisson}=0.42-0.44$ for $\alpha=-\frac{1}{2}$ \cite{Gonzalez2}.

Our considerations require the exponent $\alpha$ to be in the
range $-\frac{1}{2}<\alpha< 1$.
For $\alpha<-\frac{1}{2}$ the mean free path $l\propto  
\varepsilon^{\alpha+1/2}$ diverges at small kinetic energies.
The carriers at the current source therefore enter the conductor
ballistically and accumulate only at a finite distance from the
injection point.
Fig. \ref{fig:fig1} indicates
that the charge density at the current source must be zero
if one insists that the
electric field vanishes.
Nagaev \cite{Nagaev1998} has shown that full shot noise,
$P=P_{\rm Poisson}$, follows for $\alpha=-\frac{3}{2}$.
Presumably, $P/P_{\rm Poisson}$ will decrease monotonically
from 1 for $\alpha=-\frac{3}{2}$ to $0.38$ for $\alpha=-\frac{1}{2}$,
but we have no theory for this range of $\alpha$'s.
For $\alpha>1$ the resistance $R$ becomes infinitely large,
because the coefficient $\bar\chi(1)$
in the
current-voltage characteristic (\ref{eq:iv}) diverges.
An intuitive understanding can be obtained 
by equating the potential
gain $\phi \sim (Dt)^{3/(2\alpha+4)}$ (acquired by diffusing close
to the current source for a time $t$)
with the increase in kinetic energy $\varepsilon$: For
$\alpha>1$ this time $t\propto
\varepsilon^{(1-\alpha)/3}$ is seen to diverge for small $\varepsilon$.
We found that the shot-noise power 
vanishes as $\alpha\to 1$. Presumably, a non-zero answer for $P$ 
would follow for $\alpha> 1$ if the non-zero thermal energy and finite
charge density at the current source is accounted for. This remains
an open problem.

Discussions with O. M. Bulashenko, T. Gonz{\'a}lez,
J.~M.~J.~van Leeuwen, and W. van Saarloos are gratefully
acknowledged.
This work was supported by the European Community
(Program for the Training and Mobility of Researchers)
and by the Dutch Science Foundation NWO/FOM.

\end{document}